\documentclass[prb,twocolumn,showpacs,superscriptaddress,floatfix, nofootinbib]{revtex4-2}
\usepackage{amsmath,amssymb,amsfonts,mathrsfs, amsbsy}
\usepackage{graphicx}
\usepackage[bookmarks=true, colorlinks=true, linkcolor=blue, urlcolor=blue, citecolor=blue, bookmarks=true, hyperindex=true]{hyperref}

\newcommand{\be}{\begin{equation}}
\newcommand{\ee}{\end{equation}}

\newcommand{\beq}{\begin{eqnarray}}
\newcommand{\eeq}{\end{eqnarray}}

\def\H1{\widehat{H}_1}

\newcommand{\ket}[1]{\left| #1 \right>}

\begin{document}

\title{Many-body localization of ${\mathbb Z}_3$ Fock parafermions}

\author{M.\,S.~Bahovadinov}
\thanks{M.S.B. and W.B. contributed equally to this work.}
\affiliation{ Physics Department, National Research University Higher School of Economics, Moscow 101000, Russia}
\affiliation{Russian Quantum Center, Skolkovo, Moscow 143025, Russia}

\author{W.~Buijsman}
\thanks{M.S.B. and W.B. contributed equally to this work.}
\affiliation{Department of Physics, Ben-Gurion University of the Negev, Beer-Sheva 84105, Israel}

\author{A.\,K.~Fedorov}
\affiliation{Russian Quantum Center, Skolkovo, Moscow 143025, Russia}
\affiliation{National University of Science and Technology ``MISIS'', Moscow 119049, Russia}

\author{V.~Gritsev}
\affiliation{Institute for Theoretical Physics Amsterdam, University of Amsterdam, P.O. Box 94485, 1090 GL Amsterdam, The Netherlands}
\affiliation{Russian Quantum Center, Skolkovo, Moscow 143025, Russia}

\author{D.\,V.~Kurlov}
\affiliation{Russian Quantum Center, Skolkovo, Moscow 143025, Russia}
\affiliation{National University of Science and Technology ``MISIS'', Moscow 119049, Russia}

\date{\today}

\begin{abstract}

We study the effects of a random magnetic field on a one-dimensional (1D) spin-1 chain with {\it correlated} nearest-neighbor $XY$ interaction. We show that this spin model can be exactly mapped onto the 1D disordered tight-binding model of ${\mathbb Z}_3$ Fock parafermions (FPFs), exotic anyonic quasiparticles that generalize usual spinless fermions. Thus, we have a peculiar case of a disordered Hamiltonian that, despite being bilinear in the creation and annihilation operators, exhibits a many-body localization (MBL) transition owing to the nontrivial statistics of FPFs. This is in sharp contrast to conventional bosonic and fermionic quadratic disordered Hamiltonians that show single-particle (Anderson) localization. We perform finite-size exact diagonalization calculations of level-spacing statistics, fractal dimensions, and entanglement entropy, and provide convincing evidence for the MBL transition at finite disorder strength.

\end{abstract}

\maketitle

\section{Introduction}

The interplay between interparticle interactions and disorder in low-dimensional quantum many-body systems has been extensively studied in recent decades~\cite{Apel, Giamarchi, Altshuler} and remains an active research area (for reviews, see e.g. Refs~\cite{ ReviewLuitz, ReviewAbanin}). Numerous studies of interacting quantum many-body systems provide a strong indication of a transition to the many-body localized (MBL) phase at a sufficiently strong disorder. In the MBL phase the eigenstate thermalization hypothesis (ETH) is violated~\cite{Pal_Huse_2010, YBLev2014, Serbyn2015, Luitz2016, De_Luca_2013, De_Luca_2014}, which leads to protection of quantum states from decoherence. Recent studies of the MBL phase also demonstrate area-law entanglement of eigenstates~\cite{Nayak_2013, Serbyn2013, Gritsev2019, Serbyn2016} and vanishing steady transport~\cite{ Berkelbach2010, Barisic2016, YBLev2015, Herbrych2017}. Some of these properties can be understood with the help of emerging quasi-local conserved charges~\cite{Papic2013, Rademaker2016, Chandran2015}. Numerically, one characterizes the MBL transition by the level-spacing statistics~\cite{KudoRn, LuitzAletRn, OganesyanHuseRn, CorentinRn, ShengRn, KhemaniRn}, participation entropies~\cite{AletDq, SantosDq}, entanglement structure of eigenstates~\cite{GrayEnt, ShengRn, Gritsev2019}, quantum correlations between the neighboring states~\cite{PinoDqKl, LuitzAletRn}, and occupation spectra of one-particle density matrices~\cite{OPDMMeissner,OPDMImura,OPDMBuijsman}. 

In the vast majority of cases, the MBL problem is extremely challenging for analytical treatments and, to a large extent, the progress in this field is driven by numerical investigations. Most commonly, the focus is on the one-dimensional (1D) systems due to the computational limitations, although recent works provide evidence of MBL in two-dimensional systems~\cite{Decker2021,Assaad2021, Clark2021}. 
Paradigmatic examples of systems studied in the context of MBL are various spin-$\frac{1}{2}$ chains and ladders~\cite{LuitzAletRn,Bahovadinov2022,Baygan2015} and interacting single- and two-component fermions, e.g. the Hubbard model~\cite{Mondaini2015,Zakrzewski2018}. However, even in one dimension the physics can be much richer as the class of available physical models is much broader. For instance, models of interacting spins with the value of spin higher than $\frac{1}{2}$ often behave in a drastically different way from their spin-$\frac{1}{2}$ counterparts, starting from the famous Haldane conjecture~\cite{Haldane1,Haldane2}. Moreover, low-dimensional systems can host exotic particles with nontrivial statistics that interpolates between the conventional bosonic and fermionic ones~\cite{Wilczek1,Wilczek2,Haldane3,Wu1991,Fendley2012}. While there is a large number of studies dedicated to both higher-spin chains and anyons in one dimension (see e.g. Refs.~\cite{Babujian1982, Affleck1987, Barber1989, White1993, Eckern1998, Feiguin2007, Gong2016}), their behavior in the presence of disorder has so far received little attention and many questions remain poorly addressed.

In this paper we make a step to fill in this gap. We study a 1D spin-1 chain with {\it correlated} nearest-neighbor $XY$ interactions in the presence of a random magnetic field. Whereas the spin-$\frac{1}{2}$ $XY$ model is dual to free spinless fermions in a lattice and exhibits single-particle (Anderson) localization in the presence of disorder, this is no longer true for spin-1 $XY$ chain. Indeed, the latter is non-free and nonintegrable even without any correlations in the interaction~\cite{Haldane2,KennedyTasaki1992,Kitazawa1996}. Performing exact diagonalization calculations, we show that in the presence of a sufficiently strong random magnetic field, the spin-1 model of $XY$ type shows evident signatures of the MBL transition. 

Further, we show that the $XY$ like spin-1 chain in a random field can be exactly mapped onto the tight-binding chain with on-site disorder. Interestingly, this tight-binding model is written in terms of the so-called ${\mathbb Z}_3$ Fock parafermions (FPFs), which are exotic quasiparticles that generalize usual spinless fermions by increasing the dimensionality of their Hilbert space~\cite{Cobanera2014,Cobanera2017}. Recently, the FPF tight-binding model in the absence of disorder was studied in Ref.~\cite{Rossini2019}, where it was shown that the model is neither free nor integrable, despite being {\it bilinear} in the creation and annihilation operators of FPFs and having no explicit interaction terms. Thus, our results can be alternatively formulated as follows: the {\it quadratic} tight-binding model of ${\mathbb Z}_3$ FPFs exhibits {\it many-body} localization (and {\it not} the single-particle one) in the presence of a random on-site potential solely due to the nontrivial FPF statistics. Low-energy properties of the generalized FPF tight-binding model that includes coherent pair hopping terms were also recently investigated~\cite{MahyaehPair2020}, as well as the effects of dissipation on the FPF tight-binding chain~\cite{Mastiukova2022}.

This paper is organized as follows: In Sec.~\ref{S_model_spin}, we present the generalized spin-1 $XY$-like model of our study and discuss its main properties. In Sec.~\ref{S_map_FPF}, we briefly review ${\mathbb Z}_n$ FPFs, discuss their relation to spin-1 operators for $n$=3, and map the spin-1 chain onto the FPF tight binding model. To characterize the MBL transition, we exploit the standard set of localization probes, presented in Sec.~\ref{S_MBL_charact}. Our numerical results are presented in Sec.~\ref{S_Num_res}, and in Sec.~\ref{S_Conc} we conclude.

\section{Spin-1 chain with correlated $XY$ interaction} 
\label{S_model_spin}

We consider a spin-1 model with correlated $XY$ interactions between the neighboring spins, defined on the chain of $L$ sites with periodic boundary conditions. The Hamiltonian reads
\be \label{H_spin}
	H = - \frac{J}{2} \sum_{j=1}^L \left( S_j^{-} e^{i \Phi_j } S_{j+1}^{+} + \text{H.c.} \right) + \sum_{j=1}^L h_j S_j^z,
\ee
where we introduced a Hermitian phase operator $\Phi_j$ that reads
\be \label{Phi_op}
	\Phi_j = \phi \left(1 - S_j^z \right),
\ee
with a real-valued constant~$\phi$.
In Eq.~(\ref{H_spin}), the coupling constant is denoted by $J$ and, in what follows, we set $J=1$, the uncorrelated random magnetic field amplitudes~$h_j$ are drawn from the uniform distribution~$[-W, W]$, and $S_j^{\alpha}$ with $\alpha \in \{ z, \pm \}$ are the spin-1 operators acting nontrivially on the $j$-th site obeying the commutation relations
\be \label{spin_1_comm_rels}
	[S_j^{+}, S_k^{-}] = 2 \delta_{jk} S_j^z, \qquad [S_j^z,S_k^{\pm}] = \pm \delta_{jk} S_j^{\pm},
\ee
with $\delta_{jk}$ being the Kronecker delta. 

Hamiltonian~(\ref{H_spin}) clearly possesses the $U(1)$ symmetry associated with the conservation of total magnetization,~${\cal M} = \sum_{j=1}^L S_j^z$. In what follows, we take $L$ to be even and restrict ourselves to the ${\cal M} = L/2$ magnetization sector. In addition, we impose the constraint 
\be
	\sum_{j=1}^L h_j = 0,
\ee	
in order to ensure that mid-spectrum eigenstates taken from different disorder realizations are located in the vicinity of one and the same energy density. Sampling random numbers subject to a fixed-sum constraint can be accomplished by using the Dirichlet-rescale algorithm~\cite{Griffin2020}. 

At zero phase $\phi = 0$, Hamiltonian~(\ref{H_spin}) is simplyx the disordered spin-1 XY
model. In the absence of disorder, it has been well studied, e.g., in Refs.~\cite{KennedyTasaki1992,Kitazawa1996,Tsukano1998,Nomura1989,Sakai1990,Ejima2015}. Recently, a generalized version of the $XY$ model in the clean limit was shown to host a set of many-body scar states~\cite{Schecter2019}. Despite its simple form, the spin-1 $XY$ chain is nonintegrable, unlike its spin-$\frac{1}{2}$ counterpart that can be mapped onto free spinless fermions after the Jordan-Wigner transformation. 
Intuitively, this drastic difference between the spin-1 and spin-$\frac{1}{2}$ cases can be understood as follows.
Let us note that the 1D spin-$s$ $XY$ model can be viewed simply as describing the hopping of bosonic particles with the ladder operators $B_j^{\dag}$ and~$B_j^{\dag}$. 
However, $B_j$ and $B_j^{\dag}$ are not the traditional bosonic operators as they are subject to the hard-core constraint 
\be \label{hard_core_constraint}
	\langle B^{\dag}_j B_j \rangle \leq 2 s.
\ee
This is where the difference between $s=\frac{1}{2}$ and $s=1$ comes into play. In the former case the constraint~(\ref{hard_core_constraint}) restricts both the occupation of a given lattice site~$j$ and the real-space {\it ordering} of particles. Thus, particles can neither meet nor exchange their positions, so that there can be no many-body effects. The situation is completely different for $s=1$ since now {\it two} hard-core bosons can occupy the same lattice site. This allows particles to exchange their positions and hence leads to an effective {\it hard-core interaction}. Therefore, the spin-1 case clearly corresponds to a genuinely many-body system. We emphasize that the same hard-core interaction can be provided by adding to the spin-$\frac{1}{2}$ $XY$ chain additional terms responsible for the next-nearest-neighbor $XY$ spin-spin interactions~\cite{Bahovadinov2022}. On the other hand, one can eliminate the effective hard-core interaction from the spin-1 $XY$ chain by including an explicit interaction term of the form $U \sum_j S_j^z (S_j^z -1 )$ with~$U \to + \infty$ \cite{note_hard_core}. This reduces many-body dynamics to the single-particle one and any finite disorder localizes all eigenstates.

In Sec.~\ref{S_Num_res}, we show that the spin-1 $XY$ model with a random magnetic field [i.e. the Hamiltonian~(\ref{H_spin}) with $\phi = 0$] indeed exhibits behavior consistent with the MBL transition (see Figure~\ref{fig:r}), unlike the spin-$\frac{1}{2}$ $XY$ model that shows single-particle localization. 
It is natural to expect that the existence of the MBL phase is not specific to $\phi=0$. Further, in Sec.~\ref{S_Num_res}, we demonstrate that for $\phi = 2\pi/3$ the Hamiltonian~(\ref{H_spin}) also exhibits strong evidences for the MBL transition. We expect that the same is true for arbitrary values~$\phi$, but in this work we focus exclusively on the point~$\phi = 2\pi/3$. As we will demonstrate in the next section, in this regime Hamiltonian~(\ref{H_spin}) acquires an interesting interpretation in terms of the so-called ${\mathbb Z}_3$ FPFs.

\section{Mapping to Fock parafermions} 
\label{S_map_FPF}

\subsection{${\mathbb Z}_n$ Fock parafermions}

We now proceed with a brief overview of ${\mathbb Z}_n$ FPFs. For a more detailed discussion, see, e.g., Ref.~\cite{Cobanera2014}, where FPFs were first introduced.
FPFs are anyonic quasiparticles that generalize usual identical fermions by enlarging the dimensionality of the Fock space. Whereas for spinless fermions the local Hilbert space on a lattice site is two dimensional, for~${\mathbb Z}_n$ FPFs it is~$n$ dimensional. Thus, introducing the FPF creation and annihilation operators,~$F_j^{\dag}$ and~$F_j$ correspondingly, one has 
\be \label{FPF_number_state}
	F_j^{\dag \; m} \ket{0} = \ket{m_j}, \qquad 0\leq m \leq n-1,
\ee
where $\ket{0}$ is a vacuum, and $\ket{m_j}$ is a Fock state with~$m$ FPFs on the~$j$ th site and nothing on the rest of the chain. In other words, one can have up to~$n-1$ identical FPFs in the same state. For a general many-particle Fock state, the action of $F_j^{\dag}$ and $F_j$ is a bit more involved~\cite{Cobanera2014},
\be \label{many_body_Fock_state_FPF_ops_action}
\begin{aligned}
	&F_j^{\dag} \ket{ \ldots, m_j, \ldots } = \omega^{- \sum_{k = 1}^{j-1} m_k } \ket{ \ldots, m_j + 1, \ldots },\\
	&F_j \ket{ \ldots, m_j, \ldots } = \omega^{\sum_{k = 1}^{j-1} m_k } \ket{ \ldots, m_j - 1, \ldots },
\end{aligned}
\ee
where $\omega$ is the $n$-th primitive root of unity,
\be \label{omega_n}
	\omega = e^{2\pi i /n}.
\ee
FPF creation and annihilation operators satisfy the following relations:
\be \label{FPF_rels1}
	F_j^{\dag \; n} = F_j^n = 0, \quad F_j^{\dag \; m} F_j^m + F_j^{n-m} F_j^{\dag \; n-m} = 1,
\ee
with~$1 \leq m \leq n-1$, whereas for operators acting on different sites, one has
\be \label{FPF_braiding_rels} 
	\!\! F_j F_k = \omega^{\text{sgn}(k-j) } F_k F_j, \quad F_j^{\dag} F_k = \omega^{- \text{sgn}(k-j) } F_k F_j^{\dag},
\ee
where $\text{sgn}(x)$ is the sign function. Thus, Eqs.~(\ref{FPF_rels1}) and~(\ref{FPF_braiding_rels}) allow one to bring any monomial in FPFs to the normal ordered form. Then, we can introduce the particle number operator
\be \label{number_op}
	N_j = \sum_{m=1}^{n-1} F_j^{\dag \; m} F_j^m,
\ee
which acts on the Fock states as 
\be
	N_j \ket{ \ldots, m_j, \ldots} = m_j \ket{ \ldots, m_j, \ldots },
\ee
and satisfies the relations
\be \label{N_F_comm}
	\left[ N_j, F_j \right] = - F_j, \qquad	[ N_j, F_j^{\dag} ] = F_j^{\dag}.
\ee
Note that for $n>2$ one has $N_j^n \neq N_j$.
One can easily check that Eqs.~(\ref{FPF_rels1})--(\ref{number_op}) reduce to the standard fermionic relations for~$n=2$. 
On the other hand, for~$n>2$, from~Eq. (\ref{FPF_braiding_rels}) one clearly sees that FPFs are anyonic-type particles with the statistical parameter~$2/n$. In this case, the factor of~$\omega^{\text{sgn}(k-j)}$ in Eq.~(\ref{FPF_braiding_rels}) comes into play and leads to significant physical consequences.

Just like the usual spinless fermions are related to the spin-$\frac{1}{2}$ Pauli operators $\sigma^x_j$ and $\sigma^z_j$ via the Jordan-Wigner transformation, ${\mathbb Z}_n$ FPF creation and annihilation operators can be mapped to the generalized ${\mathbb Z}_n$ Pauli operators $X_j$ and $Z_j$ (also called shift and clock operators) via the Fradkin-Kadanoff transformation~\cite{Fradkin1980}:
\be \label{FPF_to_spin}
	F_j = \prod_{k=1}^{j-1} Z_k \, B_j, 
\ee
where $Z_k$ is the ${\mathbb Z}_n$ generalization of the Pauli operator~$\sigma_j^z$ and the operator $B_j$ is given by
\be \label{Zn_Weil_boson}
	B_j = \frac{n-1}{n} X_j - \frac{1}{n}X_j \sum_{m=1}^{n-1} Z_j^m,
\ee
with $X_j$ being the ${\mathbb Z}_n$ generalization of the Pauli operator~$\sigma_j^x$. 
The operators $X_j$ and $Z_j$ are unitary, 
\be
	X_j^{\dag} = X_j^{-1}, \qquad Z_j^{\dag} = Z_j^{-1}
\ee
and satisfy the relations
\be
	X_j^n = Z_j^n = 1, \qquad X_j^{\dag} = X_j^{n-1} , \qquad Z_j^{\dag} = Z_j^{n-1}.
\ee 
These operators commute on different sites, $X_j Z_k = Z_ k X_j$ for $k\neq j$, whereas on the same site they obey the commutation relation
\be \label{X_Z_properties}
	X_j Z_j = \omega Z_j X_j,
\ee
with~$\omega$ given by Eq.~(\ref{omega_n}). Using Eq.~(\ref{FPF_to_spin}), we see that the number operator~(\ref{number_op}) can be written simply as
\be \label{num_op_B}
	N_j = \sum_{m=1}^{n-1} B_j^{\dag \; m} B_j^m,
\ee
where we took into account that $Z_k$ is unitary and for $k\neq j$ it commutes with both $Z_j$ and $X_j$. 
Finally, let us also mention two useful identities
\be \label{exp_id}
\begin{aligned}
	& Z_j = \omega^{N_j} = e^{\frac{2\pi}{n} i N_j}, \\
	& e^{\tau N_j} = 1 + (e^{\tau} -1 ) \sum_{m=1}^{n-1} e^{(m-1)\tau} F_j^{\dag \; m} F_j^m,
\end{aligned}
\ee
where $\tau$ is an arbitrary complex number. One can easily derive Eq.~(\ref{exp_id}), e.g., using the explicit matrix representations of FPF operators given in the Appendix.

\subsection{Spin-1 chain in terms of ${\mathbb Z}_3$ Fock parafermions}

Let us now rewrite the Hamiltonian~(\ref{H_spin}) of the spin-1 disordered chain with correlated $XY$ interactions in terms of FPFs. Using the results of the Appendix and the previous section for the case of $n = 3$, we immediately obtain
\be \label{FK_transform_Z3}
	F_j = \frac{1}{\sqrt{2}} \prod_{k=1}^{j - 1} Z_k \, S_j^{+}, \qquad N_j = 1 - S_j^z.
\ee
Therefore, the spin chain Hamiltonian~(\ref{H_spin}) in terms of ${\mathbb Z}_3$ FPFs reads
\be \label{H_FPF_gen}
	H = - \sum_{j =1}^L \left( F_j^{\dag} e^{i\phi N_j}Z_j^{\dag} F_{j+1} + \text{H.c.} \right) - \sum_{j=1}^L h_j N_j,
\ee
where we used the fact that the operator (\ref{Phi_op}) becomes $\Phi_j = \phi N_j$. Then, taking into account Eq.~(\ref{exp_id}) with $n=3$, we immediately see that for $\phi = 2 \pi / 3$ the exponential operator in Eq.~(\ref{H_FPF_gen}) cancels out with the factor of $Z_j^{\dag}$. The remaining Hamiltonian is then given by
\be \label{H_FPF_tb}
  H^{\prime} = - \sum_{j=1}^L \left( F_j^{\dag} F_{j+1} + F_{j+1}^{\dag} F_j + h_j N_j \right),
\ee
which is simply the tight-binding model of ${\mathbb Z}_3$ FPFs with an onsite potential disorder.

The tight-binding ${\mathbb Z}_3$ FPF Hamiltonian~(\ref{H_FPF_tb}) with $h_j \equiv 0$ has been extensively studied in Ref.~\cite{Rossini2019}, and its generalized version that includes coherent pair hopping terms was investigated in Ref.~\cite{MahyaehPair2020}. It was shown that due to nontrivial commutation relations~(\ref{FPF_braiding_rels}) of the operators $F_j^\dagger$ and $F_j$, the FPF tight-binding Hamiltonian is nonintegrable despite being quadratic in FPF operators. In the absence of disorder, the Hamiltonian~(\ref{H_FPF_tb}) anticommutes with the parity operator $P = \exp(i \pi N_\text{even})$, where $N_\text{even}$ is the number of excitations on even lattice sites. This leads to the spectrum being symmetric around zero energy, and results in the occurrence of zero-energy eigenstates. 

We finish this section by noting that from a practical perspective it is more convenient to rewrite the Hamiltonian~(\ref{H_FPF_tb}) in terms of the operators $B_j$ and $Z_j$ since their commutation relations are much simpler than those for $F_j$ and $F_j^{\dag}$. One then has
\be \label{H_FPF_tb_2}
  H^{\prime} = - \sum_{j = 1}^L \left( B_j^{\dag} Z_j B_{j+1} + Z_j^{\dag} B_j B_{j+1}^{\dag} + h_j N_j \right),
\ee
where the number operator $N_j$ is in the form of Eq.~(\ref{num_op_B}).
Using the results of the Appendix, we immediately see that the Hamiltonian~(\ref{H_FPF_tb_2}) has complex matrix elements, which indicates that the time-reversal symmetry is broken. In the rest of the paper, we study the Hamiltonian~(\ref{H_FPF_tb_2}) and demonstrate strong evidence of the MBL transition at sufficiently strong disorder. 

\section{MBL characterizations}
\label{S_MBL_charact}

One of the contrasting signatures of the MBL phase is the absence of quantum correlations between neighbouring many-body eigenstates. This implies vanishing level repulsion, which is directly manifested in level-spacing statistics. We first analyze level repulsion in a given energy shell and then characterize quantum correlations using the Kullback - Leibler (KL) divergence~\cite{KL1}.
  
An important quantity to probe level repulsion in the many-body spectrum of the disordered model is the ratio of the minimum to maximum neighbouring minigaps,
\begin{equation}
 r_i =\frac{\min (\delta_i,\delta_{i+1})}{\max (\delta_i,\delta_{i+1})},
\qquad
\delta_i = E_i-E_{i-1},
\label{eq:r}
\end{equation}
where $E_i$ are the ordered energy eigenvalues. In the ergodic phase the level spacing distribution obeys Wigner's surmise of the Gaussian unitary ensemble (GUE). On the other hand, in the MBL phase, level repulsion is expected to vanish and there is a Poissonian distribution (PS) of the level spacings $\delta_i$. As a result, the disorder-averaged value of $\langle r \rangle$ for the Wigner-Dyson (WD) distribution is $\langle r\rangle_{\rm W} \approx 0.600$. In the localized phase, this value is $\langle r\rangle_{\rm P}=2\ln 2-1 \approx 0.386$. \citep{OganesyanHuseRn}. Thus, the crossover from the chaotic value $\langle r\rangle_{\rm W}$ to $\langle r\rangle_{\rm P}$ when disorder strength is increased captures the MBL transition.

 To quantify the quantum correlations we consider eigenstates in the occupation (Fock) basis. For the model~(\ref{H_FPF_tb}) with $L$ sites and a fixed number of parafermions ${\cal M}$ with at most $n-1 = 2$ parafermions on each sites. If one introduces the function ${\cal{P}}_z(q)$ that counts restricted partitions of a positive integer $q$ with $z$ being the maximal allowed summand, then one has ${\cal N_H}$-dimensional Hilbert space, with ${\cal N_H}=\sum_{k=1}^{{\cal{P}}_{(n-1)}({\cal M})}C^L_{k}$, where $C_k^L$ is the binomial coefficient.
 We analyze many-body eigenstates in this basis $|{\cal{S}} \rangle= |s_1 \rangle \otimes |s_2\rangle \otimes ... \otimes |s_L \rangle$, with local states $|s_i\rangle \in \left\lbrace | 0 \rangle, |1 \rangle, ..., |n-1 \rangle \right\rbrace$. 
The KL divergence is then given as KL~\cite{KL1,KL2,PinoDqKl},
\begin{equation} 
KL=\sum_{s=1}^{\cal N_H} | \psi_{\gamma}(s) |^2 \ln \left( \frac{|\psi_{\gamma}(s)|^2}{|\psi_{\gamma+1}(s)|^2} \right),
\label{eq:KL}
\end{equation} 
where the eigenstates $\ket{\psi_\gamma}$ are also ordered in energy. In the MBL phase spatial correlations of neighbouring states vanish and the ratio $| \frac{\psi_{\gamma}(s)}{\psi_{\gamma+1}(s) }|$ is exponentially large if $|\psi_{\gamma}(s)|$ is not sufficiently small. On the other hand, in the thermal phase states $|\psi_{\gamma}(s)|$ and $|\psi_{\gamma+1}(s)|$ are strongly correlated with $| \frac{\psi_{\gamma}(s)}{\psi_{\gamma+1}(s) }| \sim O(1)$ and $KL \sim O(1)$ also. Thus, one can equally identify the onset of the MBL phase with increasing disorder strength when an abrupt change of KL occurs. 

Localization in the MBL phase is equally manifested in the entanglement structure of eigenstates. We use the von Neumann entanglement entropy (EE) to quantify the entanglement between two subsystems ($A$ and $B$) after bipartiting the chain in a left- and right-hand side:
\begin{equation} 
 S_{vN} = - \text{Tr} \rho_A\ln{\rho_A},
\end{equation} 
where $\rho_A$ is the reduced density matrix for subsystem $A$. For localized states EE($L_A$) saturates with the size of subsystem $L_A$ to a constant value, when $L_A > l$, where $l$ is a characteristic correlation length. For thermal states such behavior is not obeyed and one should have the volume law for EE. Previous numerical experiments indeed demonstrated that the EE obeys area law in the MBL phase~\cite{Nayak_2013, GrayEnt}, whereas in the thermal phase a volume law is satisfied. 

\begin{figure}[t]
\includegraphics[width=\columnwidth]{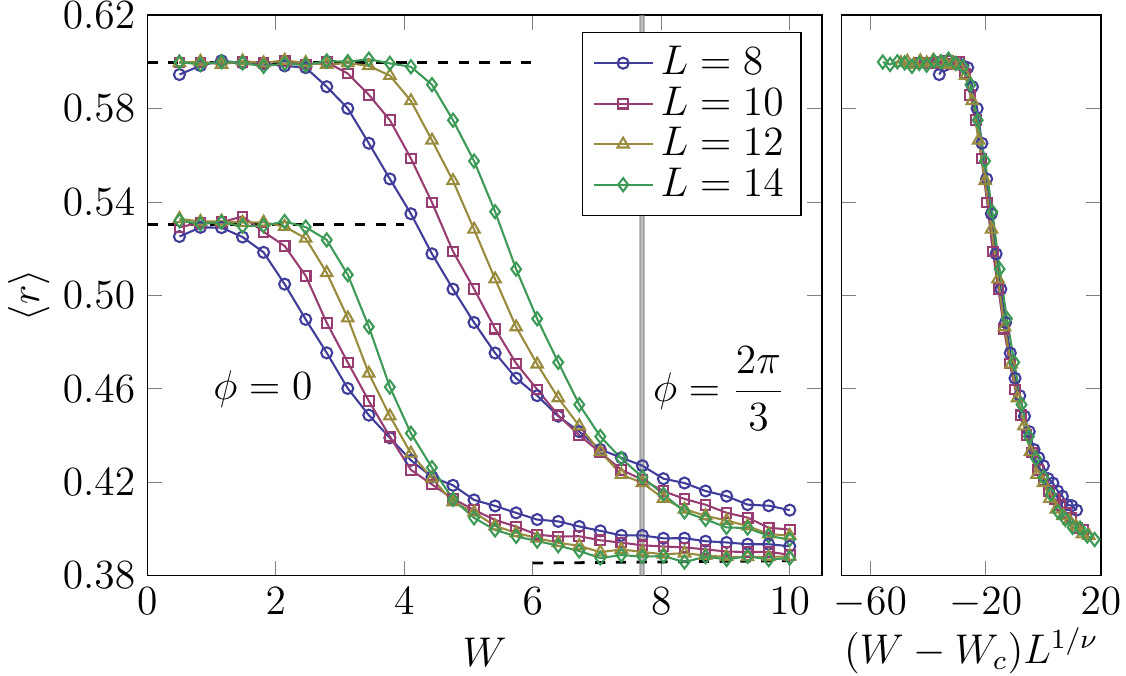}
\caption{Left panel: The average ratio of the minimum to maximum neighbouring minigaps as a function of disorder strength for the Hamiltonian (\ref{H_FPF_gen}) with $\phi = 0$ and $\phi = 2 \pi / 3$ represented by, respectively, the upper and lower sets of curves at several system sizes. Statistical errors are smaller than the marker sizes. The expectation values $\langle r \rangle \approx 0.536$, $\langle r \rangle \approx 0.600$, and $\langle r \rangle \approx 0.386$ fo,r respectively, GOE, GUE, and Poissonian level statistics are indicated by dashed lines. The vertical bar indicates the critical disorder strength for $\phi = 2 \pi / 3$ found through the finite-size scaling collapse displayed in the right panel. Right panel: A finite-size scaling collapse of the data from the left panel for $\phi= 2 \pi / 3$ with fitted critical parameters $W_c = 7.70 \pm 0.03$ and $\nu = 1.29 \pm 0.02$. See the main text for details.}
\label{fig:r}
\end{figure}

Equipartition of the many-body eigenstates $\ket{\psi_\gamma}$ in the Hilbert space is a necessary condition for thermal behavior. In the MBL phase, this is violated. We next characterize this absence of thermalization in the computational basis. Fractal dimensions $D_q$ usually serve as a standard tool for this purpose. The set of $D_q$ is determined from the scaling of participation entropies $S_q$ with~${\cal N_H}$,
\begin{equation}
S_q=\frac{1}{q-1}\ln\left(\sum_{s=1}^{\cal N_H} |\psi_\gamma(s)|^{2q}\right) \xrightarrow{{\cal N_H} \rightarrow \infty} D_q \ln\left({\cal N_H}\right) .
\label{eq:Dq}
\end{equation}
Localized eigenstates $\ket{\psi_\gamma}$ have a finite support set of $|s \rangle$ and $S_q$'s do not scale with ${\cal N_H}$. The latter implies $D_q=0$ for any $q>0$. On the other hand, thermal states satisfying ETH with $|\psi_\alpha(s)|^2 \sim {\cal N_H}^{-1} $ have $D_q=1$. The multifractal states with $0<D_q<1$ are non-ergodic albeit extended. In this study, we confine ourselves to the second fractal dimension $D_2$. 
 
To ease the numerical treatment of the model (\ref{H_FPF_gen}), we mapped the Hamiltonian onto a system of bosons using the Fradkin-Kadanoff transformation (see Sec. \ref{S_map_FPF}). We exploit shift-invert exact diagonalization to obtain a finite window of eigenstates at zero energy for system sizes ${\cal L}\in \lbrace 8,10,12,14 \rbrace$. Due to the broken time-reversal symmetry, the memory requirement to store the corresponding sparse matrices is doubled, compared to the time-reversal symmetric case. The Hilbert space dimension corresponding to the largest system size is ${\cal N_H}= 45,474$. For all system sizes, the total number of states in the considered ensemble is always less than $1\%$ of the total spectrum and the number of disorder realizations is $N_D \sim 10^3$. 
 
 \begin{figure}[t]
\includegraphics[width=\columnwidth]{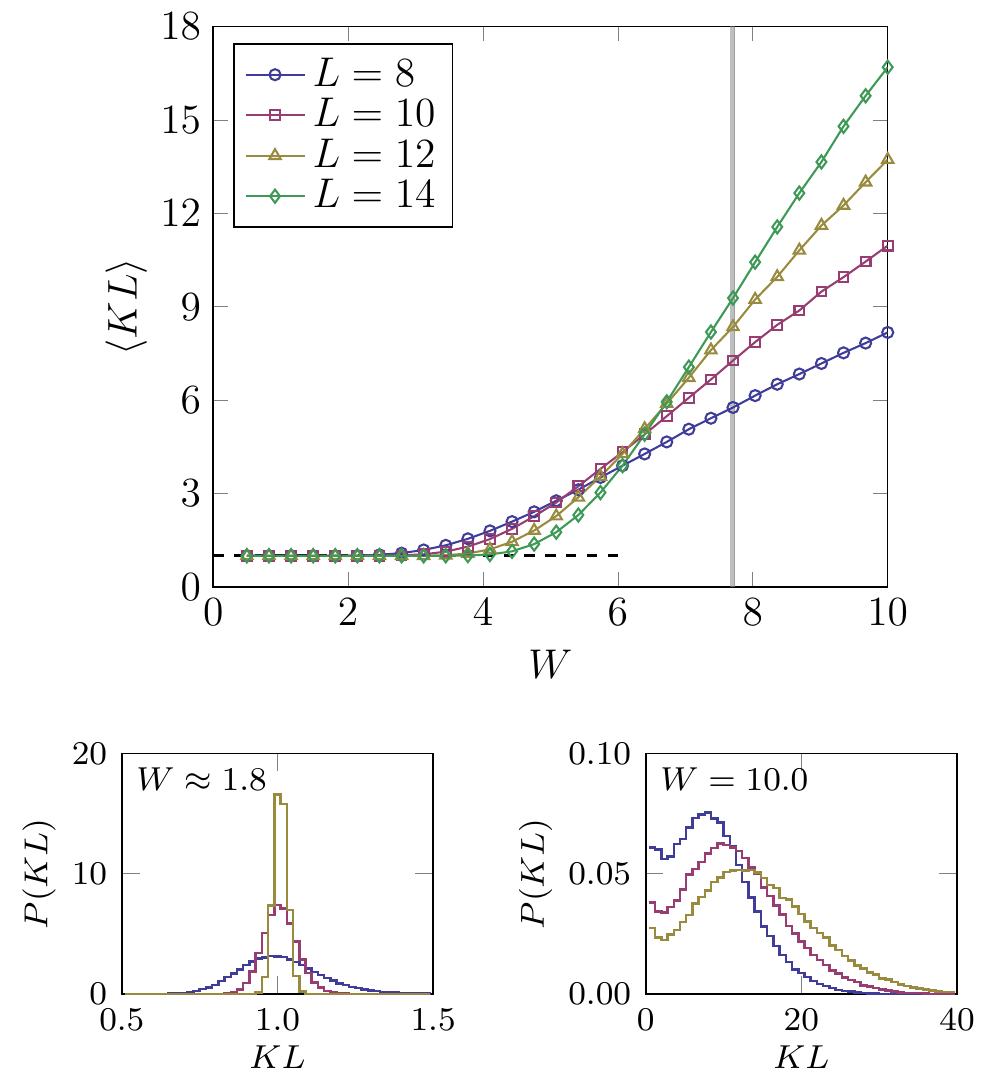}
\caption{Top panel: The average KL divergence for $\phi = 2 \pi / 3$ as a function of disorder strength at several system sizes. The expectation value $\langle KL \rangle = 1$ for ergodic systems is indicated by a dashed horizontal line. The vertical line indicates the critical disorder strength found by the finite-size scaling collapse of the average ratio of the minimum to maximum neighbouring minigaps. Statistical errors are smaller than the marker sizes. Lower left panel: The distribution of the KL divergence in the ergodic phase for several system sizes. Lower right panel: The distribution of the KL divergence in the many-body localized for several system sizes.}
\label{fig:KL}
\end{figure}

\section{Numerical results}
\label{S_Num_res}

Here, we present numerical evidence pointing towards the presence of the MBL transition in the model described by the Hamiltonian of Eq. (\ref{H_FPF_gen}). In all figures, statistical errors are smaller than the marker sizes, and are thus omitted. 

First, we focus on the eigenvalue statistics through the average ratio of the minimum to maximum neighbouring minigaps of Eq. (\ref{eq:r}). Figure \ref{fig:r} shows the ensemble average of this quantity as a function of the disorder strength at several system sizes for $\phi=0$ and $\phi = 2\pi / 3$. At weak disorder strengths, the GUE (Gaussian orthogonal ensemble [GOE)]) $\langle r \rangle \approx 0.600$ ($\langle r \rangle \approx 0.536$ for $\phi=0$) for ergodic systems is approached. For strong disorder, $\langle r \rangle$ tend towards the value for Poissonian level statistics $\langle r \rangle \approx 0.386$. In the present context, this indicates localization of the many-particle system. Rescaling $W \to (W - W_c) L^{1/\nu}$ with numerically optimized parameters $W_c = 7.70 \pm 0.03$ and $\nu = 1.29 \pm 0.02$ results in a universal curve, plotted in the right panel of Figure \ref{fig:r}. Uncertainty estimates for the critical parameters have been obtained by bootstrap analysis. The present finite-size scaling analysis suggests the presence of the MBL transition in the thermodynamic limit for both $\phi=0$ and $\phi=2\pi/3$ and hence supports our qualitative picture presented in the previous sections. Hereafter, we limit ourselves solely to the $\phi = 2 \pi / 3$ case and expect qualitatively similar results for other values of $\phi \neq 0$ (broken time-reversal symmetry).
 
 \begin{figure}[t]
\includegraphics[width=\columnwidth]{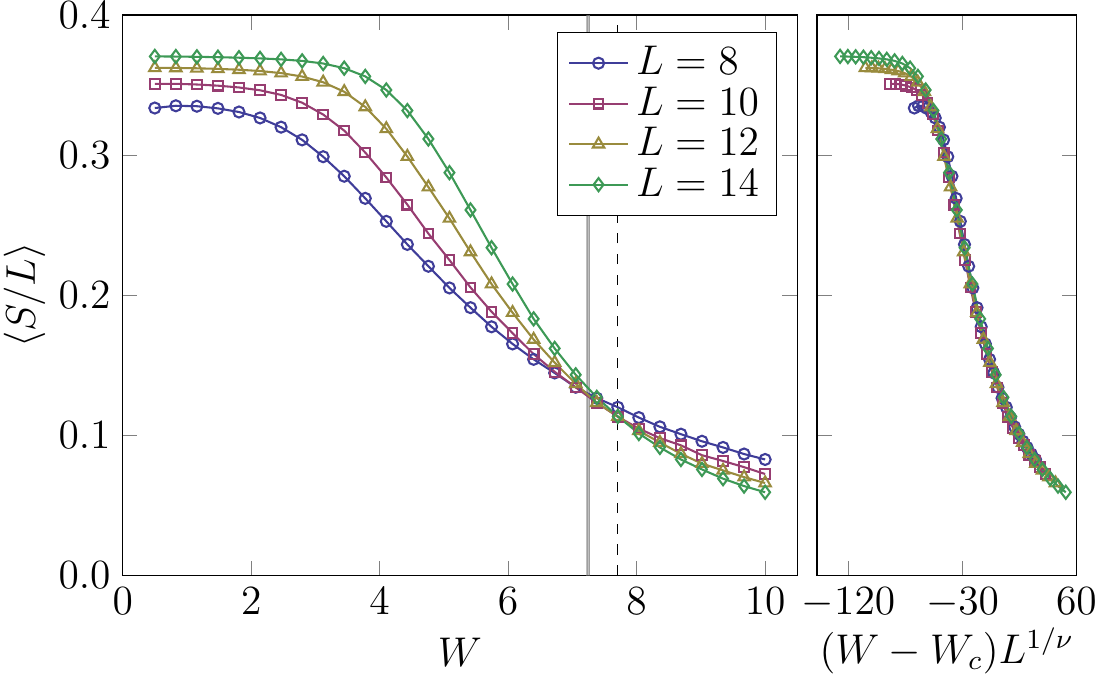}
\caption{Left panel: The average bipartite entanglement entropy per site for $\phi = 2 \pi / 3$ as a function of disorder strength at several system sizes. The system is divided into equally sized left- and right-hand sides. Statistical errors are smaller than the marker sizes. The vertical dotted line indicates the critical disorder strength found through the finite-size scaling collapse for the ratio of the minimum to maximum neighbouring minigaps. The vertical bar indicates the critical disorder strength found through the finite-size scaling collapse displayed in the right panel. Right panel: A finite-size scaling collapse of the data from the left panel with fitted critical parameters $W_c = 7.25 \pm 0.02$ and $\nu = 0.90 \pm 0.0049$. See the main text for details.}
\label{fig:EE}
\end{figure}

Next, we focus on the statistical properties of the eigenstates. First, we consider the KL divergence $KL$ of Eq. (\ref{eq:KL}). Figure \ref{fig:KL} shows the ensemble average of the KL divergence for the eigenstates as a function of disorder strength at several system sizes. Ergodic and correlated eigenstates are characterized by $\langle KL \rangle = 1$, which is observed at low disorder strengths. In the thermodynamic limit, localized systems are characterized by an extensive KL divergence. The crossing points of the curves increases with increasing system size rather significantly. As such, no quantitative estimate for the critical disorder strength can be obtained. However, focusing on the distribution of the KL divergence in the lower panels of the figure, we observe two qualitatively different phases at low ($W \approx 1.8$) and high ($W=10.0$) disorder strength. In the thermal phase, the width of the Gaussian distribution of KL shrinks with system size, exhibiting self-averaging behavior. On the other hand, in the MBL phase this feature is lost.

Next, we focus on the average eigenstate entanglement entropy per lattice site. In the ergodic phase, the eigenstate entanglement entropy obeys volume-law scaling. Figure \ref{fig:EE} shows the average eigenstate entanglement entropy per site for a decomposition of the system in equally sized left- and right-hand sides. The data indicates that the system is in an ergodic phase at low disorder strengths. Localized systems obey area-law scaling of the eigenstate entanglement entropy. In the thermodynamic limit, one thus expects $\langle S / L \rangle \to 0$. A finite-size scaling collapse similar to the one for the average ratio of the minimum to maximum neighbouring minigaps indicates a transition at the critical disorder strength $W_c = 7.25 \pm 0.02$, where the uncertainty indicates the statistical error. Because of smaller sample-to-sample variations than above, the error is smaller here. With increasing system size, the scaling tends towards a volume (an area) law for $W< W_c$ ($W > W_c$). Following the discussion above, this provides supporting evidence for a many-body localization transition at a disorder strength close to the one found above.

\begin{figure}[t]
\includegraphics[width=\columnwidth]{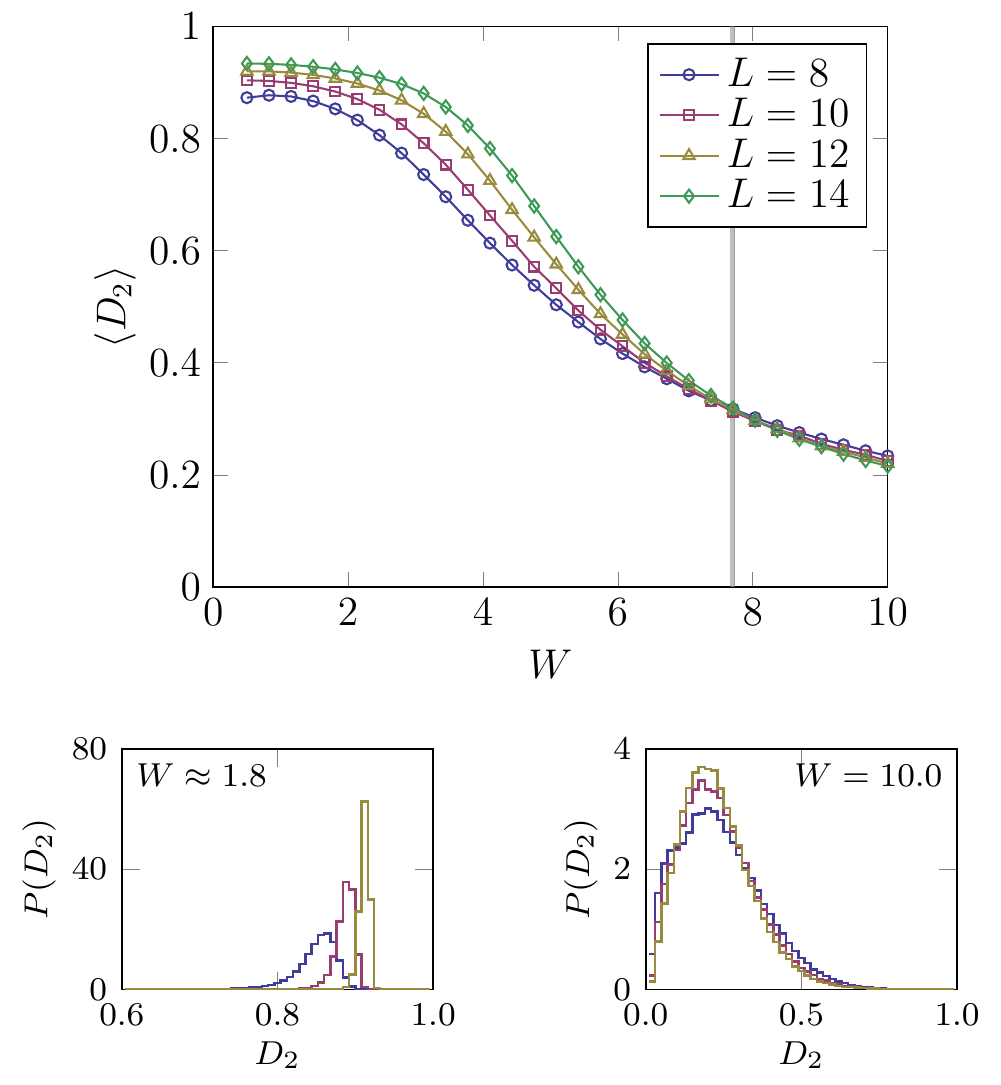}
\caption{Main panel: The average second fractal dimension as a function of disorder strength for $\phi = 2 \pi / 3$ at several system sizes. The vertical line indicates the critical disorder strength found by the finite-size scaling collapse of the average ratio of the minimum to maximum neighbouring minigaps. Statistical errors are smaller than the marker sizes. Lower left panel: The distribution of the second fractal dimension in the ergodic phase for several system sizes. Lower right panel: The distribution of the second fractal dimension in the many-body localized for several system sizes.}
\label{fig:D2}
\end{figure}

Finally, we consider the average second fractal dimension $\langle D_2 \rangle$. Ergodic systems are characterized by $D_2 = 1$ (eigenstates are spread out over the full Hilbert space), while localized systems obey $\langle D_2 \rangle = 0$ (eigenstates are spread out over a vanishing fraction of the Hilbert space). Figure \ref{fig:D2} shows the average of the second fractal dimension as a function of disorder strength at several system sizes. At low disorder strength, the curves tend towards $D_2 = 1$ with increasing system size, while they tend to $D_2 = 0$ at strong disorder strength. The crossing points of the curves are close to the critical disorder strength for the average ratio of the minimum to maximum neighbouring minigaps. Focusing on the distribution of $D_2$ in the lower panels, we observe that the ergodic and many-body localized phase are qualitatively different: the distribution of $D_2$ in the ergodic phases has vanishing moments with the system size, whereas in the localized phase it broadens and shows opposite skewness.

\section{Conclusions}
\label{S_Conc}

In this work, we provide numerical evidences for the MBL transition in a disordered tight-binding chain of ${\mathbb Z}_3$ Fock parafermions and the dual correlated spin-1 $XY$ model. At weak disorder, the system is in an ergodic phase and eigenstate thermalization hypotheses is obeyed. This is shown by the GUE/GOE statistics of energy minigaps and strong level correlations, quantified via KL calculations. At strong disorder, level correlations vanish and hence Poisson statistics of energy minigaps is exhibited. These distinctive features at weak and strong disorder strongly suggest that the system undergoes the MBL transition at intermediate disorder strength. These results are supported by quantified Hilbert space equipartition and entanglement of many-body eigenstates at small/large disorder strengths, characterized by fractal dimensions and entanglement entropy. 

Let us conclude with a few remarks regarding the nature of the transition, found in this work. It is well known (see, e.g., Ref.~\cite{Evers2008}) that one of the key differences between the Anderson (single-particle) and the MBL transitions in one-dimensional systems is their occurrence at arbitrarily small and finite disorder strength, correspondingly. In other words, observing a transition to the localized phase at finite disorder strength (which is the case in this work) provides strong evidence for a genuine MBL transition. 
Thus, the disordered Fock parafermionic tight-binding model~(\ref{H_FPF_tb}) provides us with an interesting example of a Hamiltonian that demonstrates a behavior consistent with MBL, despite being quadratic in terms of creation/annihilation operators of particles with a well-defined Fock space. This observation is in striking difference with more conventional one-dimensional fermionic or bosonic disordered bilinear Hamiltonians, which exhibit Anderson localization.
On the other hand, the presence of an MBL transition in the Hamiltonian~(\ref{H_FPF_tb}) is quite natural, since in the absence of disorder the model cannot be described in terms of free particles~\cite{Rossini2019}. In fact, to the best of our knowledge, all known (Fock) parafermionic models that allow for a single-particle description are necessarily non-Hermitian~\cite{Fendley2014, Mastiukova2022}. From this perspective, it would also be very interesting to find a Hermitian parafermionic model that exhibits a single-particle localization. We leave this question for future work.

\begin{acknowledgements}
 We thank Boris Altshuler and Dirk Schuricht for useful discussions. This research was supported in part through computational resources of the HPC facilities at HSE University~\cite{Kostenetskiy_2021}. The work of VG is part of the DeltaITP consortium, a program of the Netherlands Organization for Scientific Research (NWO) funded by the Dutch Ministry of Education, Culture and Science (OCW). WB acknowledges support from the Kreitman School of Advanced Graduate Studies at Ben-Gurion University. MSB thanks Basic Research Program of HSE for the provided support. AKF thanks the support of the RSF grant (19-71-10092) and the Priority 2030 program at the National University of Science and Technology ``MISIS” under the project K1-2022-027 (analysis of the method).

{\it Note added.} While finishing the manuscript we have learned that the authors of Ref.~\cite{Camacho2022} obtained similar results on MBL of ${\mathbb Z}_3$ FPFs. As far as our studies overlap, our results are in agreement with each other. 
\end{acknowledgements}

\appendix

\section{Matrix representation of ${\mathbb Z}_n$ operators}
For completeness, in this Appendix we provide matrix representations of~${\mathbb Z}_n$ operators~$X_j$,~$Z_j$, and $B_j$.
The so-called shift $X_j$ and clock $Z_j$ operators act nontrivially on $j$th site and their matrix representations are $X_j = 1\otimes \ldots 1 \otimes X \otimes1 \ldots$ and $Z_j = 1\otimes \ldots \otimes Z \otimes \ldots$, where $X$ and $Z$ are~$n \times n$ matrices given by
\be
	X = \begin{pmatrix}
	0&1 & 0 & \ldots & 0\\
	0 & 0 & 1 & \ldots & 0\\
	 & \vdots & & \ddots &\\
	0 & 0 & 0 &\ldots &1 \\
	1& 0 & 0 & \ldots & 0
	\end{pmatrix}, \quad 
	Z = \begin{pmatrix}
		1 & 0 & \ldots & 0 \\
		0 & \omega & \ldots & 0 \\
		\vdots & & \ddots & \vdots \\
		0& 0 & \ldots & \omega^{n-1}
		\end{pmatrix}.
\ee
In compact form, the matrix elements of $X$ and $Z$ are $X_{a,b} = \delta_{(a \text{ mod } n) + 1, b}$ and $Z_{a,b} = \delta_{a,b} \omega^{a-1}$, correspondingly. Clearly, for~$n =2$ one has~$\omega = - 1$ and the shift and clock matrices reduce respectively to the Pauli matrices: $X = \sigma^x$, $Z = \sigma^z$. Then, from Eq.~(\ref{Zn_Weil_boson}) for the operators $B_j$ we have the representation $B_j = 1\otimes \ldots \otimes B \otimes \ldots$ with the $n \times n$ matrix $B$ given by
\be
	B = \begin{pmatrix}
	0&1 & 0 & \ldots & 0\\
	0 & 0 & 1 & \ldots & 0\\
	 & \vdots & & \ddots &\\
	0 & 0 & 0 &\ldots &1 \\
	0& 0 & 0 & \ldots & 0
	\end{pmatrix}.
\ee
Likewise, one has $N_j = 1\otimes \ldots \otimes N \otimes \ldots$ with a diagonal matrix $N = \text{diag}\{ 0, 1 , 2, \ldots, n - 1 \}$. 

Then, restricting ourselves to the case~$n=3$ and taking into account that spin-$1$ operators from Eq.~(\ref{spin_1_comm_rels}) are represented as
\be
	S^z = \begin{pmatrix}
	1 & 0 & 0\\
	0 & 0 & 0 \\
	0 & 0 & -1
	\end{pmatrix}, \quad
	S^+ = \sqrt{2} \begin{pmatrix}
	0 & 1 & 0\\
	0 & 0 & 1 \\
	0 & 0 & 0
	\end{pmatrix},
\ee
we immediately obtain $S^+ = \sqrt{2} B$ and $S^z = 1 - N$, which gives us Eq.~(\ref{FK_transform_Z3}) in the main text.


\begin{thebibliography}{99}
\bibitem{note_hard_core}
    In terms of FPFs the corresponding term reads $U \sum_{j} N_j (N_j - 1)$, where $N_j$ is the number operator given by Eq.~(\ref{number_op}).
\end{thebibliography}
\end{document}